Variational B-Rep Model Analysis

**Variational B-rep Model Analysis for Direct Modeling using Geometric Perturbation**

Qiang Zou, Hsi-Yung Feng.
Department of Mechanical Engineering
The University of British Columbia
Vancouver, BC
Canada V6T 1Z4

**Abstract**

The very recent CAD paradigm of direct modeling gives rise to the need of processing 3D geometric constraint systems defined on boundary representation (B-rep) models. The major issue of processing such variational B-rep models (in the STEP format) is that free motions of a well-constrained model involve more than just rigid-body motions. The fundamental difficulty lies in having a systematic description of what pattern these free motions follow. This paper proposes a geometric perturbation method to study these free motions. This method is a generalization of the witness method, allowing it to directly deal with variational B-rep models represented with the standard STEP scheme. This generalization is essentially achieved by using a direct, geometric representation of the free motions, and then expressing the free motions in terms of composites of several basis motions. To demonstrate the effectiveness of the proposed method, a series of comparisons and case studies are presented.

**1. Introduction**

Computer-aided design (CAD) is a tool widely used in mechanical design. Direct modeling is the very recent CAD modeling paradigm that features direct manipulation of the geometry of the model to much increase model editing flexibility [1,2]. With this technique, models can be easily edited through intuitive operations [3]. Two directions may be taken to implement direct modeling: integrating direct modeling into history-based modeling [4], or integrating variational modeling [5] into direct modeling.

The former integration direction stated above is conceptually simple but has inherent limitations. The integration is often done by simply adding direct edits to the end of the model construction history as additional feature operations, and the original history remains exactly as before. As pointed out by [6], this kind of integration could mess up the parametric information in a model and lead to the loss of meaningful parametric controls. One possible solution to this limitation is to develop a new module translating direct edits into operations of tuning parameters of the features already presented in the model construction history. This way of working may help but cannot solve the limitation altogether because not all direct edits are achievable through feature parameter tuning. To make matters worse, feature parameter tuning for a given direct edit (if achievable) is often not unique. In this regard, this work opts for the latter integration direction that integrates variational modeling into direct modeling.

To integrate variational modeling into direct modeling, an additional information layer of geometric constraints is added on top of the boundary representation (B-rep) model used in direct modeling. Direct edits only change the geometry (i.e., the B-rep layer) of the model. As a result, the geometry-constraint consistency in the pre-edit model would be broken, and the constraint layer of the model needs to be adapted accordingly. That is, the geometric constraint system (GCS) in the constraint layer is to be changed by updating the constraint parameters with the dimensions on the altered model geometry and removing any inapplicable constraints. This update process could break the well-constrained state of the pre-update GCS, resulting in an under-constrained and/or over-constrained model. It is thus necessary to develop methods to analyze the constrained state of the updated GCS, and then according to the outcome, to add/remove constraints such that the GCS becomes well-constrained again. The objective of this work is to provide an automatic GCS analysis method, which can assist the user to effectively

---

. Corresponding author. Tel.: +1-604-822-1366; fax: +1-604-822-2403. E-mail: feng@mech.ubc.ca

Variational B-Rep Model Analysis

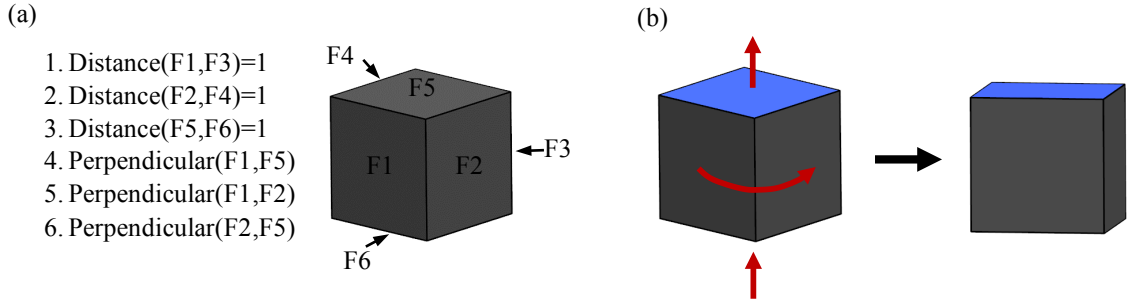

(a)
1. Distance(F1,F3)=1
2. Distance(F2,F4)=1
3. Distance(F5,F6)=1
4. Perpendicular(F1,F5)
5. Perpendicular(F1,F2)
6. Perpendicular(F2,F5)

(b)

Figure 1   Example of a well-constrained variational B-rep model (a) and a non-rigid-body free motion (b).

handle the non-well-constrained model after direct edits while not requiring him/her to become an expert in the mathematics of geometric constraint reasoning.

GCS analysis has been extensively studied in the geometric constraint solving domain [6]. Among these studies, the witness method [7,8] is the most suitable method for the analysis task of this work. However, a direct application of the witness method is insufficient and new developments are necessary. The witness method [8] assumes a condition of no redundancy on geometry representation parameters. This condition can be understood with the help of the following example: a plane may be represented with (1) a tuple $(a, b, c, d)$ and an additional constraint $a^2 + b^2 + c^2 = 1$ (the plane equation is $ax + by + cz + d = 0$), and (2) a point $p \in R^3$ on the plane and the normal $n \in S^2$ of the plane. Scheme (1) has no redundant representation parameters, while Scheme (2) has redundant representation parameters because $p$ is not unique and any point on the plane can be used to define it. The above condition for applying the witness method has not been discussed in previous studies. We will validate this statement in the Implementation and results section. This condition is not necessarily impossible to satisfy, but the standard B-rep scheme (i.e., the STEP standard [9]) used in CAD is not of such a quality. This issue might be solved by using two different representation schemes in parallel in CAD, which however introduces another issue: an additional CAD module translating between the schemes is needed. This work opts for a direct method that can handle variational B-rep models with the standard STEP scheme. Hereafter, variational B-rep models refer to those with the standard STEP scheme, unless stated otherwise.

The challenge of having a direct method is that free motions (without violating the GCS) of a well-constrained model involve more than just rigid-body motions. Consider, for example, the well-constrained model in Fig. 1a. Rigid-body motions are clearly free motions. Some non-rigid-body motions are also applicable. By translating the top and bottom planes and rotating the four side planes in a way as in Fig. 1b, we attain a motion that is not a rigid-body motion, and meanwhile does not violate the model's GCS. A systematic study of this kind of free motions must be made in order to have an effective analysis of variational B-rep models. To handle this challenge, a geometric perturbation method is to be presented, showing that there is an elegant pattern behind these free motions: any such motion can be expressed in terms of a composite of several basis motions. The presented method essentially generalizes the witness method so that it can directly handle variational B-rep models with the standard STEP scheme.

## 2. Related work

In the last three decades, a good number of publications related to GCS analysis have been presented, refer to [6] for a thorough review of them. The ideas presented can be roughly classified into four categories: solving-based, logic-based, graph-based and perturbation-based. Among them, the solving-based methods are the (conceptually) simplest, which analyze the constrained state of a GCS by directly solving the GCS through either symbolic methods (e.g., Grobner bases, Wu-Ritt triangulation methods) or numerical methods (e.g., Newton-Raphson, homotopy continuation methods) [6]. These methods are rarely used in GCS analysis practices due to the high computational load.

The logic-based methods like [10,11] develop a set of geometric theorems and derivation rules and then, based on them, check if the GCS can be logically derived. If so, the GCS is well-constrained; if there are extra constraints, it is over-constrained; otherwise, it is under-constrained. This process is essentially an axiomatization of all possible constraint systems, which is of great mathematical elegance. Yet, the library of the geometric theorems and derivation rules developed is far from being complete for practical usages.

The graph-based methods handle the GCS in an indirect way by converting a GCS to a graph structure and then studying this graph instead of the original GCS. There are two lines of development. The first one tries to recognize pre-set graph patterns (corresponding to known shapes) in a GCS. This line was pioneered by Owen [12] and much improved by [13–15] in the size of the pattern library. The second line compares degrees of freedom (DOFs) of the geometry with degrees of restriction of the GCS. This line was initialized by Bardford [16] and Serrano [17], and detailed by [18–20]. In 2001, the two lines were summarized under a unified framework by Hoffmann et al. [21]. Ever since, there have still been some good progress, e.g., [22], but the foundations remain unchanged. Although graph-based methods could be effective in many scenarios, a known limitation is the incapability to handle constraint dependencies (except for the simplest structural dependencies) [7]. The reason is that the graph representation only captures the combinatorial information of the GCS and discards the geometric information.

To overcome the above limitation of graph-based methods, the witness method was proposed [7]. This method examines how the constraint equations behave under the infinitesimal perturbations made to the equation variables. The relationship between the behavior and perturbations is described by the associated Jacobian matrix of the constraint equations; the dependent rows of this matrix characterize the (structural and non-structural) constraint dependencies and the kernel of this matrix gives the DOFs of the GCS [8,23,24]. A successful application of this method requires the Jacobian to be evaluated at a carefully selected point called witness configuration, which is not straightforward to generate in the context of geometric constraint solving [25,26]. However, this task becomes trivial in direct modeling: the witness configuration is simply the current model geometry (i.e., the model geometry after the direct edits). This geometry has satisfied all the model constraints due to the pre-processing of GCS update that synchronizes the model GCS with the model geometry. In this regard, the witness method fits the variational B-rep model analysis task very well. To adopt this approach, new developments are however needed to resolve the challenge as stated in Section 1.

### 3. Methodology

#### 3.1. Problem statement

**Variational B-rep model.** A variational B-rep model consists of a B-rep model and a GCS defined on top of the B-rep model. The data in a B-rep model are usually classified as geometry (i.e., the geometric entities including surfaces, curves, and points) and topology (i.e., the connections between these geometric entities) [27]. What a constraint in the model GCS actually references are the geometric entities in the B-rep model. A geometric entity is usually represented by several parameters, and then a constraint can be described by one or more algebraic equations defined over these parameters. Thus, the model GCS can be described by a system of nonlinear equations $F(X) = 0$, where $X$ is the vector of geometry parameters (viewed as equation variables). Here, we assume that the equations $F(X) = 0$ are independent of the used Cartesian frame.

**Definition 1.** Let $F(X) = 0$ be the constraint equations of a GCS, and $S$ the solution space. The GCS is consistently constrained if $S \neq \emptyset$, and inconsistently constrained otherwise.

**Definition 2.** Let $F(X) = 0$ represent a consistently constrained GCS, and $S$ the solution space. The GCS is under-constrained if the cardinality $|S|$ is infinite (two solutions are considered identical if their points sets are equal modulo a displacement).

**Definition 3.** Let $F(X) = 0$ represent a consistently constrained GCS, and $S$ the solution space. The GCS is consistently over-constrained if there exists a subset of the GCS such that the corresponding reduced equations have the same solution space as $S$.

**Definition 4.** A GCS is over-constrained if it is inconsistently constrained or consistently over-constrained.

**Definition 5.** A GCS is well-constrained if it is neither under-constrained nor over-constrained.

**Definition 6.** A model is rigid if its associated GCS is consistently constrained and not under-constrained.

**Problem statement.** Given a variational B-rep model (the GCS has already been updated), determine its constrained state and then, if not well-constrained, group constraints that form minimal over-constrained subsystems and cluster geometric entities that form maximal rigid subsystems.

#### 3.2. The proposed geometric perturbation method

Let $F(X) = 0$ be the constraint equations of the given GCS, where $X$ denotes the equation variables. The witness method applies a small disturbance $\Delta X$ to $X$ and examines the response $\Delta F$ of $F$, which are related by:

$$\Delta F = J(X) \cdot \Delta X + O(\|\Delta X\|^2) \tag{1}$$

where $J(X)$ is the Jacobian matrix evaluated at $X$. As mentioned, the witness configuration undergoing perturbation in direct modeling is the model geometry after direct edits, and thus the $X$ here is to be drawn from this geometry. If there exist dependent constraints, their responses to the perturbations should be related, or equivalently the rows of the Jacobian matrix that correspond to these constraints are dependent. Then, over-constraint is described by the vectors in the kernel (or the null space) of $J^T$. Under-constraint is closely related to the free perturbations that do not violate existing constraints, i.e., $\Delta F = J \cdot \Delta X = 0$. That is, under-constraint is described by the vectors in the kernel of $J$.

The perturbations that the witness method [7,8] applies are parametric, as in Eq. 2. The free perturbations given by this method are thus not able to show how the model flexes in terms of geometric motions (translations and/or rotations) explicitly. Geometric motions are not only beneficial for designers to understand under-constraint but also important for studying the non-rigid-body free motions as in Fig. 1. Foufou and Michelucci [8] and Haller et al. [28] hardcoded the parametric-to-geometric conversion for linear geometries such as lines and planes. However, other commonly used geometries like cylinders were not included. This subsection presents a generic method for applying perturbations in terms of geometric motions (abbreviated as geometric perturbations).

A geometric motion of a geometric entity consists of two time-varying components: $R(t)$ to specify the rotational part of the motion and $T(t)$ to describe the translational part. At time $t$, the geometric entity's position vector $p(t)$ and orientation vector $n(t)$ are expressed as:

$$p(t) = R(t)p(0) + T(t)$$
$$n(t) = R(t)n(0)$$
(2)

Differentiating this equation gives the relationship between the parametric perturbations and geometric perturbations [29]:

$$\delta p = \delta r \times p(t) + \delta t$$
$$\delta n = \delta r \times n(t)$$
(3)

where $\delta r$ is the vector describing the rotation about the direction $\delta r / \|\delta r\|$ by the angle $\|\delta r\|$, and $\delta t$ the vector describing the translation. Denoting $p(t)$ by $(p_x,\ p_y,\ p_z)^T$ and $n(t)$ by $(n_x,\ n_y,\ n_z)^T$, the cross products $\delta r \times p(t)$ and $\delta r \times n(t)$ in Eq. 3 can be rewritten in the following form:

$$\delta r \times p(t) = -\begin{pmatrix} 0 & -p_z & p_y \\ p_z & 0 & -p_x \\ -p_y & p_x & 0 \end{pmatrix}\delta r = -P\delta r$$

$$\delta r \times n(t) = -\begin{pmatrix} 0 & -n_z & n_y \\ n_z & 0 & -n_x \\ -n_y & n_x & 0 \end{pmatrix}\delta r = -N\delta r$$
(4)

Eventually, Eq. 3 can be rewritten as a linear transformation:

$$\begin{pmatrix} \delta p \\ \delta n \end{pmatrix} = \begin{pmatrix} I_{3\times3} & -P \\ 0 & -N \end{pmatrix}\begin{pmatrix} \delta t \\ \delta r \end{pmatrix}$$
(5)

This equation describes the parametric-to-geometric conversion for a single geometric entity. To attain the conversion for all the geometric entities in the model, we just need to assemble each entity's conversion matrix diagonally as follows:

$$\begin{pmatrix} \delta p_1 \\ \delta n_1 \\ \vdots \\ \delta p_m \\ \delta n_m \end{pmatrix} = \begin{pmatrix} T_1 & & & \\ & T_2 & & \\ & & \ddots & \\ & & & T_m \end{pmatrix}\begin{pmatrix} \delta t_1 \\ \delta r_1 \\ \vdots \\ \delta t_m \\ \delta r_m \end{pmatrix}$$
(6)

where $T_i = \begin{pmatrix} I_{3\times3} & -P_i \\ 0 & -N_i \end{pmatrix}$. By denoting the transformation matrix in Eq. 6 as $T$ and multiplying it with the Jacobian matrix $J$ in Eq. 1, the new matrix $G = JT$ relates any geometric perturbations made to the GCS and its responses. That is, this matrix allows the witness method to apply geometric perturbations directly to a GCS. Hereafter the matrix $G$ is referred to as the geometric perturbation matrix.

It should be noted here that the transformation matrix formulated in Eq. 6 only reaches the parametric changes of the geometric entities' positions and orientations, which is enough for the standard representation scheme as it represents a geometric entity with its type, position, orientation and size dimensions (if any). In fact, the transformation matrix in Eq. 6 can be readily extended to dealing with an arbitrary representation scheme. What we need is an additional conversion from the changes of the positions and orientations to those of the specific representation parameters; such a conversion is straightforward and can even be done automatically (Chapter 8, [30]). Let $T'$ denote this conversion matrix. The geometric perturbation matrix as stated above is then given as the product of the three matrices in the following form $G = JT'T$, where $J$ is the Jacobian matrix in Eq. 1 and $T$ the transformation matrix in Eq. 6.

### 3.3. GCS characterization

A well-constrained GCS has a finite number of solutions or feasible geometric configurations. Since these feasible configurations are finite, one feasible configuration cannot be continuously deformed to another without violating the existing constraints. That being said, no free motions of a well-constrained GCS will alter the model's shape. It is proven in Proposition 1 that there are in total three types of geometric motions that preserve model shape: (1) rigid-body motions; (2) geometry-invariant motions; and (3) composites of them. The geometry-invariant motions for a geometric entity are motions that leave the entity invariant (e.g., translating a plane along any direction in the plane). Fig. 2 illustrates the geometry-invariant motions of commonly used geometric entities in CAD, including lines, planes, cylinders, spheres, cones, and tori. With Proposition 1, the solution-based characterization of a well-constrained GCS is translated to the conditions on the free motions of the GCS.

| Geometric entities | Invariant motions | Illustrations |
|---|---|---|
| Line | 1. Translations along the line<br>2. Rotation about the line |  |
| Plane | 1. Translations in the plane<br>2. Rotations about the normal |  |
| Cylinder | 1. Translations along the axis<br>2. Rotations about the axis |  |
| Sphere | 1. Any rotations |  |
| Cone | 1. Rotations about the axis |  |
| Torus | 1. Rotations about the axis |  |

Figure 2  Illustrations of geometry-invariant motions (vertex, i.e., sphere with radius 0, is the same as sphere; ellipsoid of revolution or surface of revolution are the same as cone).

**Proposition 1.** *Free motions of a well-constrained GCS are either rigid-body motions or geometry-invariant motions, or composites of them.*

**Proof.** See Appendix.

**Remark.** The concept of geometry-invariant motions may seem superficial if we only consider a single geometric entity. It is however important when a bunch of geometric entities is involved. That is, what really matters is the composite outcome of these motions. Such composites significantly complicate the free motions of a well-constrained variational B-rep model. For example, the free motion described in Fig. 1b is a composite of (1) a translational geometry-invariant motion for planes F1, F2, F3 and F4, and (2) a rotational geometry-invariant motion for planes F5 and F6.

Before presenting the details of GCS characterization, several terms are made precise. *Free motions* are motions that do not violate the constraints in the GCS. *Rigid-body motions* are motions that translate and/or rotate the model as a whole. *Geometry-* *invariant motions* have been defined in the above paragraph. *Flexion motions* are free motions that are neither rigid-body motions nor geometry-invariant motions (intuitively, flexion motions deform the model).

As already discussed, a well-constrained GCS does not have any free motions other than rigid-body and geometry-invariant motions. In other words, the well-constrained state of a GCS can be determined by comparing the set of free motions with the set of rigid-body and geometry-invariant motions. Owing to the linear feature of the proposed geometric perturbation method, the set of free motions has a simple representation: the linear subspace spanned by the free motion vectors, and the same for the representation of the set of rigid-body and geometry-invariant motions. The comparison of the two sets is also made easy: checking the difference between the two spanned linear spaces. More specifically, the free motion space is given by the kernel of the geometric perturbation matrix $G$:

$$Ker(G) = \{x | Gx = 0\} \tag{7}$$

The space of rigid-body and geometry-invariant motions is given by the image of the matrix whose columns are the vectors representing rigid-body motions and geometry-invariant motions. Denote this matrix as $B_{m \times n}$. Its image is defined as [31]:

$$Im(B) = \{Bx | x \in R^n\} \tag{8}$$

Next, we show how to generate the vectors of rigid-body and geometry-invariant motions.

The rigid-body motions have six motion bases: three axial translations and three axial rotations. The translation along a coordinate axis, e.g., the $x$-axis, is given by:

$$\left( \underbrace{1,0,0}_{\delta t_1}, \underbrace{0,0,0}_{\delta r_1} \quad \cdots \quad \underbrace{1,0,0}_{\delta t_m}, \underbrace{0,0,0}_{\delta r_m} \right)^T \tag{9}$$

where $m$ denotes the number of geometric entities. The representation of axial rotations is similar. For example, the rotation about the $x$-axis is:

$$\left( \underbrace{0,0,0}_{\delta t_1}, \underbrace{1,0,0}_{\delta r_1} \quad \cdots \quad \underbrace{0,0,0}_{\delta t_m}, \underbrace{1,0,0}_{\delta r_m} \right)^T \tag{10}$$

The geometry-invariant motions are translational/rotational motions along/about some specific vectors (Fig. 2). Let the vector be $v = (v_x, v_y, v_z)^T$. A translational geometry-invariant motion along this vector for the $k$-th geometric entity is:

$$\left( \underbrace{0,0,0}_{\delta t_1}, \underbrace{0,0,0}_{\delta r_1} \quad \cdots \quad \underbrace{v_x, v_y, v_z}_{\delta t_k}, \underbrace{0,0,0}_{\delta r_k} \quad \cdots \quad \underbrace{0,0,0}_{\delta t_m}, \underbrace{0,0,0}_{\delta r_m} \right)^T \tag{11}$$

and the same for a rotational geometry-invariant motion about the vector:

$$\left( \underbrace{0,0,0}_{\delta t_1}, \underbrace{0,0,0}_{\delta r_1} \quad \cdots \quad \underbrace{0,0,0}_{\delta t_k}, \underbrace{v_x, v_y, v_z}_{\delta r_k} \quad \cdots \quad \underbrace{0,0,0}_{\delta t_m}, \underbrace{0,0,0}_{\delta r_m} \right)^T \tag{12}$$

The vectors in Eq. 9-12 establish a set of basis vectors for the space of rigid-body and geometry-invariant motions.

By comparing the two spaces $Ker(G)$ and $Im(B)$ (refer to Eq. 7 and 8), whether the GCS (without constraint dependencies) is well-constrained or not can be determined. Alternatively, noticing that the space of free motions always includes the space of rigid-body and geometry-invariant motions, we just need to compare the dimensions of the two spaces: $Dim\big(Ker(G)\big) - Dim\big(Im(B)\big) = ColumnSize(G) - Rank(G) - Rank(B)$[1], where the equivalence is due to the rank-nullity theorem [31], and $Dim(\cdot)$ is an operator getting the dimension of a linear space. The difference $Dim\big(Ker(G)\big) - Dim\big(Im(B)\big)$ is called the degree of flexion of the GCS in this work and denoted as $DFLX(G)$. Then, the criterion for being well-constrained is:

$$DFLX(G) = 0 \tag{13}$$

where $Dim(\cdot)$ gets the dimension of a linear space. If Eq. 13 does not hold, i.e.,

$$ColumnSize(G) - Rank(G) - Rank(B) > 0 \tag{14}$$

---

[1] Remark: $Dim(\cdot)$ is for linear space, and $Rank(\cdot)$ is for linear mapping.

the GCS is under-constrained. In the above discussion, it is assumed that there are no constraint dependencies. If such dependencies occur, the perturbation matrix will have linearly dependent rows, i.e., $Ker(G^T) \neq \emptyset$ [8], which is equivalent to:

$$Rank(G) - RowSize(G) < 0 \tag{15}$$

The three criteria in Eq. 13-15 yield a generalization of those provided in the original witness method. They are able to characterize the free motions as stated in Section 1. Additionally, they show that there is an elegant, linear structure behind these free motions.

### 3.4. GCS decomposition

By using the criteria developed in Section 3.3, whether a given GCS is well-constrained or not can be determined. If it is well-constrained, nothing further needs to be done. If otherwise, the task of finding minimal dependent subsystems and maximal rigid subsystems will be activated.

As mentioned, constraint dependencies are encoded as linear dependencies among rows of the geometric perturbation matrix $G$. To find the sets of the minimal dependent rows in $G$, a $QR$ decomposition based method is to be used. This method is a variant of the methods presented in [8,24]. Their methods are incrementally; the method presented here works in a unified, one-time manner. The $QR$ decomposition of the matrix $G^T$ with column-pivoting gives three matrices $P$, $Q$ and $R$ such that [30]:

$$G^T P = QR \tag{16}$$

Here, $P$ is a permutation matrix, $Q$ a unitary matrix and $R$ an upper triangular matrix. The first $k = Rank(G^T)$ columns of the matrix $G^T P$ make up a maximal set of independent columns. Here, $Rank(G^T)$ can be readily given by the $QR$ decomposition. Then, the matrix $G^T P$ can be divided into two submatrices:

$$G^T P = \begin{bmatrix} \underset{first\ k\ columns}{C} & \underset{rest\ columns}{D} \end{bmatrix} \tag{17}$$

The dependency of any column vector $d \in D$ with the column vectors in $C$ can be expressed in terms of solving the linear system $Cx = d$. Since $C$ is full column rank, there is a unique solution to this linear system and the involved columns are minimal. Extending Eq. 17 to including all the columns of $D$ yields:

$$CX = D \tag{18}$$

This system can also be readily solved using the $QR$ decomposition results [30]. Eq. 18 can be rewritten as:

$$\begin{bmatrix} C & -D \end{bmatrix} \begin{bmatrix} X \\ I_{n \times n} \end{bmatrix} = 0 \tag{19}$$

where $n = ColumnSize(D)$. A column vector in $[X\ I_{n \times n}]^T$ corresponds to a dependency group and the non-zero elements of this vector indicate the involved columns in $G^T P$. Finally, to get the dependency among the original constraints, we need to map the columns of $G^T P$ to those of $G^T$. This is done by multiplying the vector in $[X\ I_{n \times n}]^T$ with the permutation matrix $P$ in Eq. 16. The resulting vector's non-zero elements give the indices of the constraints involved in the dependency group.

Maximal rigid subsystems are important for understanding under-constraint. As mentioned, under-constraint is described by the flexion motions of the GCS, which are encoded in vectors in the kernel of the geometric perturbation matrix: $Ker(G)$. However, a vector $x \in Ker(G)$ does not show rigid subsystems explicitly. A subdivision method (Algorithm 1) is thus used to generate maximal rigid subsystems. In each recursion of the subdivision (Lines 2-10), the method focuses on one given free motion vector, and divides a set of geometric entities into two groups such that any two geometric entities from the respective groups move as a flexion under the given free motion vector. In the division, we do not allow any shared geometric entities between the two groups. This subdivision procedure will be applied to all the free motion vectors in a recursive manner (Lines 11-14). In Line 13, the union of two partitions is defined as follows: a partition is a set of maximal rigid subsystems and a subsystem is a set of geometric entities; the union of two partitions yields a new set collecting all the subsystems in the two partitions. The function $isFlexion(P_1, g, v)$ (Line 6) is the key to this algorithm, which checks if the given free motion $v$ is a flexion motion for the geometric entities $P_1 \cup \{g\}$. This checking can be done by solving a simple linear system. Let the free motion of the involved geometric entities be $(\delta t_i, \delta r_i)^T, i \in P_1 \cup \{g\}$. Cascading the vectors $\{(\delta t_i, \delta r_i)^T\}$ gives the column vector $b$. The construction of the motion bases for rigid-body and geometry-invariant motions is same to Eq. 9-12 but with the geothe metric entity number $m$ being reduced to the cardinality of $P_1 \cup \{g\}$. Let the bases be represented by $B$. Then, the checking is to solve the following linear system:

$$Bx = b \tag{20}$$

If this equation is unsolvable, the given free motion is not a rigid-body motion or a geometry-invariant motion for $P_1 \cup \{g\}$, i.e., it is a flexion motion, and thus the function $isFlexion(P_1, g, v)$ returns true; otherwise, it returns false.

Algorithm 1 aims to compute a feasible partition, not the best one. Consider a trivial example: there are 3 points $A, B, C$, constrained with specified distance $AB$ and $AC$. Clearly, two partitions are possible: ($AB$ and $C$) or ($AC$ and $B$). In such situations, Algorithm 1 can only give one feasible partition, depending on the initialization. Generating all possible partitions and determining which one of them is the best in terms of understanding under-constraint would be an interesting research topic for its own sake. This work, in the current form, is not able to deal with such tasks.

| **Algorithm 1**: subdivideSystem($v, G$) |
|---|
| **Input:** $v, G$ − a free motion and a set of geometric entities |
| **Output:** $P$ − subdivision results of $G$ |
| 1.  $P \leftarrow \emptyset$  // an array storing the partitions of the model |
| 2.  **while** $G \neq \emptyset$ do |
| 3.      $g_1 \leftarrow$ popFrontElement($G$) |
| 4.      $P_1 \leftarrow \{g_1\}, P_2 \leftarrow \emptyset$ |
| 5.      **for** each geometric entity $g \in G$ **do** |
| 6.          **if** $isFlexion(P_1, g, v)$ **then** |
| 7.              $P_2 \leftarrow P_2 \cup \{g\}$ |
| 8.          **else** $P_1 \leftarrow P_1 \cup \{g\}$ |
| 9.      **end for** |
| 10.     $G \leftarrow P_2$ |
| 11.     $v_{next} \leftarrow$ getNextFreeMotion() |
| 12.     **if** $v_{next} \neq NULL$ **then** |
| 13.         $P \leftarrow P \cup$ subdivideSystem($v_{next}, P_1$) |
| 14.     **else** $P \leftarrow P \cup \{P_1\}$ |
| 15. **end while** |
| 16. **Return** $P$ |

## 4. Implementation and results

This section gives the implementation details of the proposed GCS analysis method and demonstrates its effectiveness. The effectiveness of the GCS characterization criteria (Section 3.3) is to be shown by comparing with the witness method [8] using two examples. The effectiveness of the GCS decomposition methods (Section 3.4) is to be shown by three case studies.

### 4.1. Implementation details

The GCS analysis methods presented previously have been implemented using C++ on top of the geometric modeling kernel OpenCASCADE (version 7.0). The implementation was embedded in a larger programming framework developed through using the geometry processing and rendering framework OpenFlipper [32] as a reference source. Direct modeling edits were carried out through the software prototype developed in the authors' previous work [3]. The main part of the proposed methods includes construction and manipulation of the geometric perturbation matrix (for example, the $QR$ decomposition in Section 3.4), solving linear systems, and comparing dimensions of linear spaces. These tasks were accomplished by using the linear algebra library Eigen (version 3.2.9). As usual in CAD/CAM, a tolerance of $10^{-7}$ is used for deciding nullity; no tolerance tuning was needed.

### 4.2. Comparison examples

**Example description.** The comparisons of the proposed GCS characterization criteria with the witness method were conducted using the two examples shown in Fig. 3. The first example (Fig. 3a) involves four planes, and the associated constraints are distance and perpendicularity between them. The second example (Fig. 3b) is based on two 3D lines, and the (only) constraint is the distance between them. These two configurations are widely used as constituent features of mechanical parts. For example, the configuration in Fig. 3a is often used as the through hole feature. Both examples are well-constrained, and thus can be used to test if the proposed method and witness method can correctly characterize the constraint states.

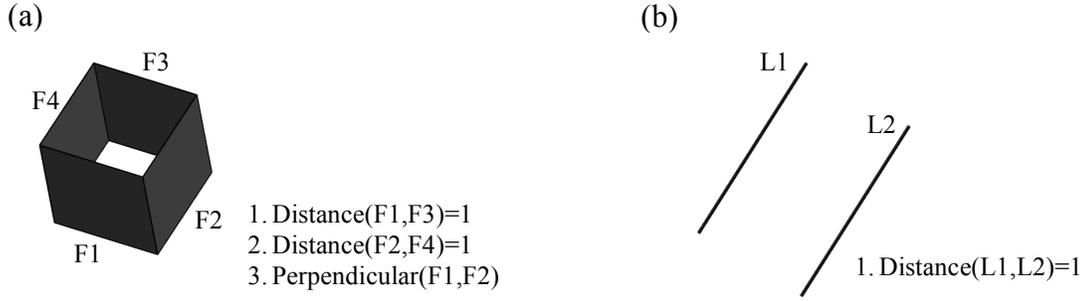

Figure 3   Comparison examples: (a) the plane example; and (b) the 3D line example.

The comparison results are given in Table 1, where the columns show characterization results for the respective methods. Each method's column is further subdivided into three sub-columns that specify, for each method, its two important intermediate results and the correctness of the final characterization result. The intermediate results include the calculated degrees of freedom (see the *Real DOF* columns) and the expected degrees of freedom (see the *Nominal DOF* columns). The values in the *Nominal DOF* columns were calculated as follows: for the witness method, they are set to the degree of displacement/rigidity [33]; and for the proposed method, they are set to $Dim(Im(B))$ as in Eq. 13. The rows of the table represent the examples to which the above results refer.

Table 1   Comparison results for the plane and line examples (DOF: degree of freedom).

| | Witness Method | | | Proposed Method | | |
|---|---|---|---|---|---|---|
| | Real DOF | Nominal DOF | Matched? | Real DOF | Nominal DOF | Matched? |
| The Plane Example (Representation #1) | 5 | 5 | ✓ | 17 | 17 | ✓ |
| The Plane Example (Representation #2) | 11 | 6 | ✗ | 17 | 17 | ✓ |
| The Line Example | 7 | 6 | ✗ | 8 | 8 | ✓ |

As can be seen from the table, the first example was used twice with different representation schemes. The purpose of doing so is to show that the correctness of the witness method depends on the used representation scheme. This will confirm our statement about the condition of no redundant parameters for applying the witness method in the Introduction section. Representation #1 defines a plane with the tuple $(a, b, c, d)$ and the additional constraint $a^2 + b^2 + c^2 = 1$ (the plane equation is $ax + by + cz + d = 0$). Representation #2 uses the standard representation scheme, which describes a plane with a point $p \in R^3$ on the plane and the normal $n \in S^2$ of the plane. The scheme used to represent the lines in the second example consists of a point $p \in R^3$ on the line and the direction $d \in S^2$ of the line. The constraint-to-equation translation for these representation schemes is detailed as follows: (1) collinearity of two normal or direction vectors $v_1, v_2$ is represented with $v_1 + tv_2 = 0$ where $t$ is a scalar unknown; and (2) point-to-point distance and vector angle are represented with dot products.

**Discussion.** From Table 1, the proposed method is seen to be more effective than the witness method, as expected. The results shown in the Witness Method column confirm our claim about the non-redundancy condition for applying the witness method in the Introduction section. It is not necessarily impossible to design a new representation scheme satisfying the non-redundancy condition; however, based on the two examples in this subsection, we can conclude that this new scheme will be very different from the STEP standard. The proposed method extended the witness method, making it able to directly deal with the standard representation scheme. In fact, this method can deal with arbitrary representation schemes, as shown by the characterization results in the first two rows of Table 1.

*4.3. Evaluation cases*

**Case description.** Three case studies have been conducted to show the effectiveness of the GCS analysis method presented in this work. The first case study considered the direct editing of a hexahedron model (Fig. 4), which altered the model's constraint state from well-constrained to over-constrained (Fig. 5). The second case study involved the direct editing of a slot model (Fig. 7), which changed the model's constraint state from well-constrained to under-constrained (Fig. 8). To be more specific, the push-pull edit moved face F9 towards the right, removing face F6, extending face F8, and shrinking face F2. These two case studies were intended to show the respective effectiveness of the two constituent modules of the proposed GCS analysis method. The third case study analyzed a comprehensive situation, wherein the updated GCS had under-, well-, and over-constrained parts. It was used to show the effectiveness of the overall method. This case study was based on a real-world mechanical part, wherein the source model is a common bracket part and the target one is a specialized bracket part used in jet engines (Fig. 10). (To simplify the problem, the fillets in the models were suppressed since they are secondary features.) The many constraints, equations, and variables involved in this case (Fig. 11) make it hard to analyze the model manually; and it is the purpose of this case study to show how the algorithms presented previously can assist the designer in this complex analysis process.

For this work to be self-contained, the representation scheme used in the cases and the associated constraint-to-equation translation are specified as follows. A plane is represented with a point $p \in R^3$ on the plane and the normal $n \in S^2$ of the plane. A cylinder is represented with the axis, which consists of a point $p \in R^3$ on the axis and the direction $d \in S^2$ of the axis, and the radius parameter. The constraint types involved in the three case studies are: plane-plane distance, plane-plane angle, plane-plane parallel, edge length, plane-cylinder distance, cylinder-cylinder coaxial, and plane-cylinder tangent. The translation for the first three types is the same to that in Section 4.2. The translation for the rest constraint types is as follows: (1) edge length: $\|v_1 - v_2\|^2 = l^2$ where $v_1, v_2$ are the end points and $l$ the length; (2) plane-cylinder distance: $n^T d = 0$, $n^T(p_p - p_c) = l$ where $(p_p, n)$ denotes the plane, $(p_c, d)$ the cylinder axis, and $l$ the distance; (3) Plane-cylinder tangent: $n^T d = 0$, $n^T(p_p - p_c) = r$ where $(p_p, n)$ denotes the plane, $(p_c, d)$ the cylinder axis, and $r$ the cylinder radius; and (4) cylinder-cylinder coaxial: $d_1 - d_2 = 0$, $d_1 \times (p_1 - p_2) = 0$ where $(p_1, d_1)$ is the first cylinder's axis and $(p_2, d_2)$ the second cylinder's axis.

(a)                                        (b)

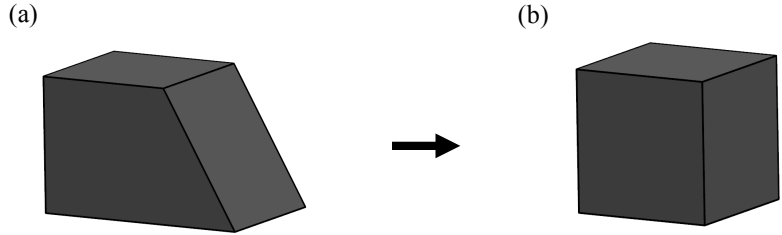

Figure 4   Push-pull direct modeling of the hexahedron model.

| Model Geometry Indices | Model GCS | | |
| --- | --- | --- | --- |
| | Original Constraints | Updated Constraints | Equations |
| | C1. Distance(F1,F3)=1 | Distance(F1,F3)=1 | 1-4 |
| | C2. Distance(F5,F6)=1 | Distance(F5,F6)=1 | 5-8 |
| | C3. Perpendicular(F1,F5) | Perpendicular(F1,F5) | 9 |
| | C4. Perpendicular(F1,F4) | Perpendicular(F1,F4) | 10 |
| | C5. Perpendicular(F4,F5) | Perpendicular(F4,F5) | 11 |
| | C6. Angle(F2,F4)=120° | Parallel(F2,F4) | 12-14 |
| | C7. Angle(F2,F5)=60° | Perpendicular(F2,F5) | 15 |
| | C8. Length(E1)=1 | Length(E1)=1 | 16 |

Figure 5   GCS udpate of the hexahedron case (Units: mm).

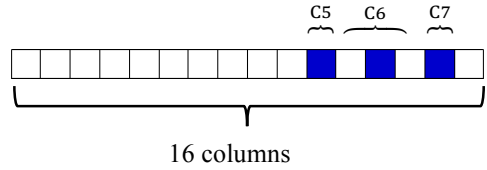

Figure 6   The dependency vector of the hexahedron case (white elements indicate the zeros and blue ones the non-zeros).

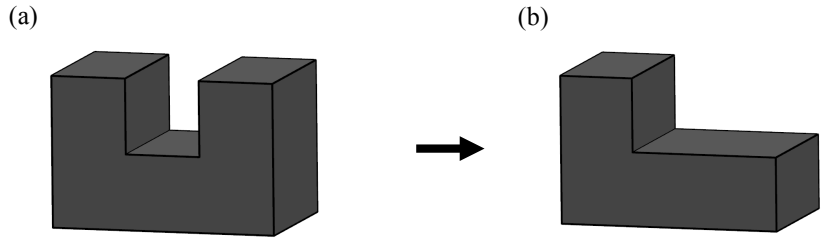

Figure 7   Push-pull direct modeling of the slot model.

| Model Geometry Indices | Model GCS | | |
|---|---|---|---|
| | Original Constraints | Updated Constraints | Equations |
| | C1.  Distance(F1,F3)=10 | Distance(F1,F3)=10 | 1-4 |
| | C2.  Distance(F2,F4)=15 | Distance(F2,F4)=15 | 5-8 |
| | C3.  Distance(F5,F10)=10 | Distance(F5,F10)=10 | 9-12 |
| | C4.  Distance(F6,F10)=10 | Deleted | N/A |
| | C5.  Perpendicular(F1,F2) | Perpendicular(F1,F2) | 13 |
| | C6.  Perpendicular(F1,F10) | Perpendicular(F1,F10) | 14 |
| | C7.  Perpendicular(F2,F10) | Perpendicular(F2,F10) | 15 |
| | C8.  Distance(F7,F9)=5 | Deleted | N/A |
| | C9.  Distance(F2,F9)=5 | Deleted | N/A |
| | C10. Distance(F5,F8)=5 | Distance(F5,F8)=5 | 16-19 |

Figure 8   GCS udpate of the slot case (Units: mm).

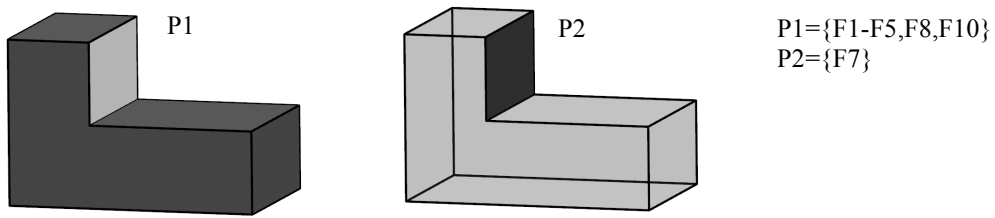

P1={F1-F5,F8,F10}
P2={F7}

Figure 9   Results of maximal rigid subsystems decomposition for the slot case.

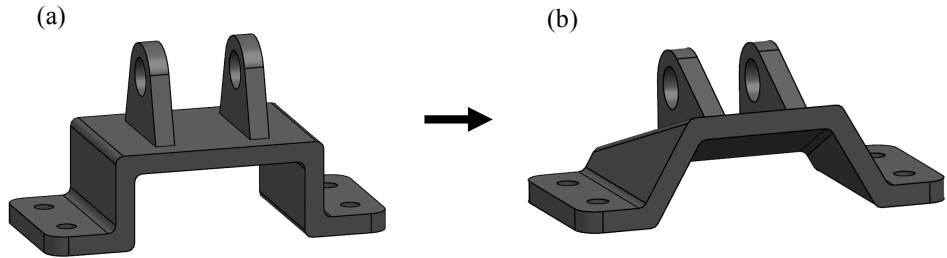

Figure 10   Push-pull direct modeling of the bracket model.

| Model Geometry Indices | Updated Model GCS | | | |
| --- | --- | --- | --- | --- |
| | Constraints | Equations | Constraints | Equations |
| | C1.  Dis(F2,F3)=10 | 1-4 | C21. Ang(F23,F13)=60° | 53 |
| | C2.  Dis(F4,F5)=10 | 5-8 | C22. Ang(F4,F22)=30° | 54 |
| | C3.  Dis(F6,F7)=10 | 9-12 | C23. Dis(F22,F24)=10 | 55-58 |
| | C4.  Dis(F8,F9)=10 | 13-16 | C24. Dis(F28,F6)=10.62 | 59-60 |
| | C5.  Dis(F10,F11)=10 | 17-20 | C25. Dis(F28,F13)=57.96 | 61-62 |
| | C6.  Ang(F1,F4)=30° | 21 | C26. Coaxial(F28,F26) | 63-68 |
| | C7.  Ang(F1,F8)=150° | 22 | C27. Tan(F28,F29) | 69-70 |
| | C8.  Dis(F1,F12)=180 | 23-26 | C28. Tan(F28,F25) | 71-72 |
| | C9.  Dis(F13,F14)=80 | 27-30 | C29. Par(F25,F14) | 73-75 |
| | C10. Ang(F3,F6)=160° | 31 | C30. Ang(F29,F13)=60° | 76 |
| | C11. Ang(F6,F11)=160° | 32 | C31. Ang(F8,F30)=30° | 77 |
| | C12. Per(F1,F3) | 33 | C32. Dis(F27,F30)=10 | 78-81 |
| | C13. Per(F1,F2) | 34 | C33. Dis(F15,F1)=10 | 82-83 |
| | C14. Per(F2,F3) | 35 | C34. Dis(F15,F13)=20 | 84-85 |
| | C15. Dis(F19,F6)=10.62 | 36-37 | C35. Dis(F16,F1)=10 | 86-87 |
| | C16. Dis(F19,F13)=57.96 | 38-39 | C36. Dis(F16,F14)=20 | 88-89 |
| | C17. Coaxial(F19,F21) | 40-45 | C37. Dis(F17,F12)=10 | 90-91 |
| | C18. Tan(F19,F20) | 46-47 | C38. Dis(F17,F13)=20 | 92-93 |
| | C19. Tan(F19,F23) | 48-49 | C39. Dis(F18,F12)=10 | 94-95 |
| | C20. Par(F20,F14) | 50-52 | C40. Dis(F18,F14)=20 | 96-97 |

Figure 11   GCS udpate of the bracket case (Dis: Distance, Ang: Angle, Per: Perpendicular, Par: Parallel, Tan: Tangent; Units: mm).

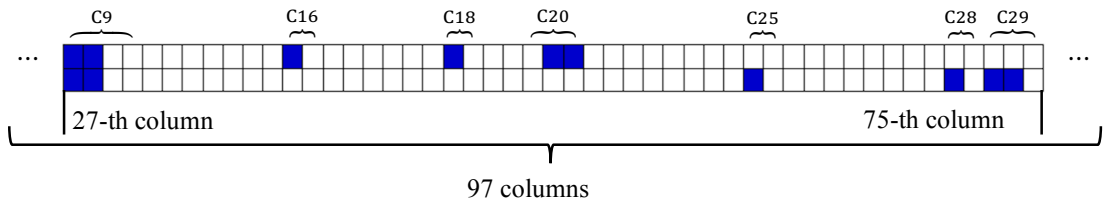

Figure 12   Truncated dependency vectors of the bracket case (truncated parts are zeros; white elements indicate the zeros and blue ones the non-zeros).

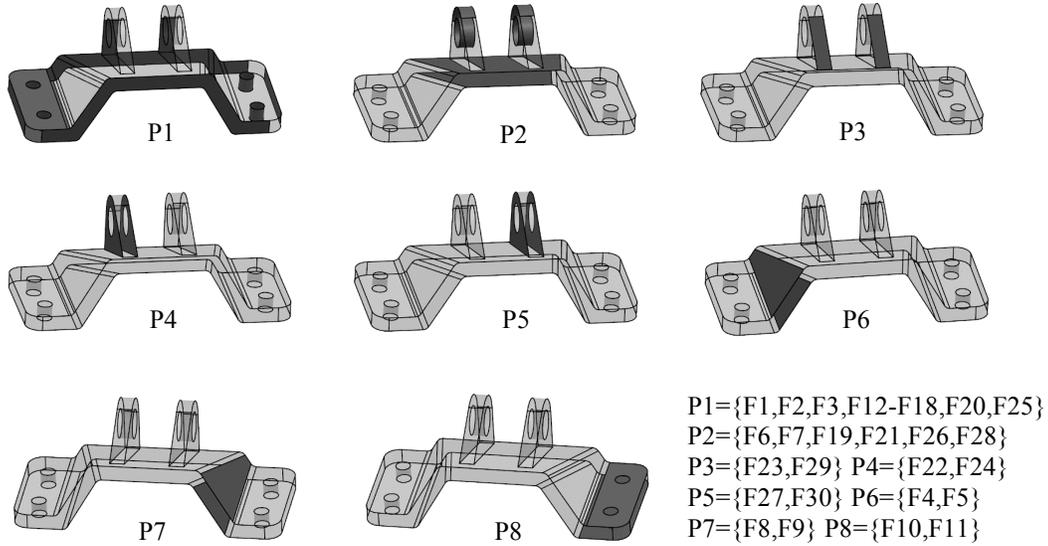

P1={F1,F2,F3,F12-F18,F20,F25}
P2={F6,F7,F19,F21,F26,F28}
P3={F23,F29} P4={F22,F24}
P5={F27,F30} P6={F4,F5}
P7={F8,F9} P8={F10,F11}

Figure 13   Results of maximal rigid subsystems decomposition for the bracket case.

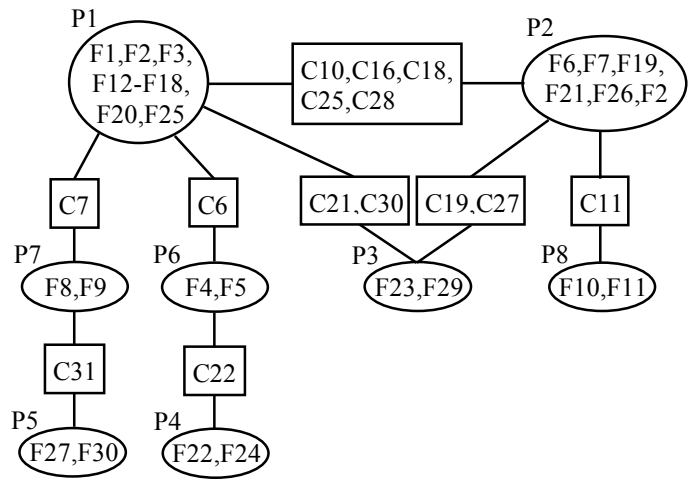

Figure 14   Bridging constraints (in rectangles) of the bracket case.

**Discussion.** The first case was purposely chosen to contain no under-constraint so that constraint dependency analysis becomes dominant in the GCS analysis. Fig. 6 shows the dependency vector given by applying the proposed decomposition method to the GCS of this case. Recall that the positions of the non-zero elements in a dependency vector indicate the constraints involved in the dependency group. Thus, the specific dependency group for this case is: {C5,C6,C7}. Attributing to the small number of constraints/equations involved in this case, the correctness of the result can be verified easily. Constraints C5, C6, and C7 form an over-constrained part because the parallel constraint C6 transitively passes the perpendicular relation (resulting from constraint C5) from the plane pair F4,F5 to another plane pair F2,F5, which is however specified again by constraint C7.

The second case has no over-constraint, and thus under-constraint analysis becomes dominant in the GCS analysis. This allows us to evaluate exclusively the maximal rigid subsystem decomposition method presented in Algorithm 1. Fig. 9 shows the results of applying Algorithm 1 to the GCS of this case. The model was divided into two parts: one consisting of planes F1-F5,F8 and F10, and the other having a single plane F7. Like in the first case, the number of constraints/equations involved in this case is still small, and the correctness of the above result can be verified manually. That is, the constraints C1-C3,C5-C7 comprise a variant of a well-constrained cuboid; constraint C10 attaches the plane F8 to the cuboid variant in a fully constrained manner; then the planes F1-F5,F8 and F10 form a well-constrained subsystem. There is no constraint between the two parts, and thus the plane F7 is free and comprises a rigid subsystem by itself.

The first and second case studies are manually manageable, which are thus not able to show fully the effectiveness of the proposed method. In this regard, the third case study analyzed a real-world mechanical model that has 97 equations in the GCS (Fig. 11). Some of them are redundant and the whole GCS is insufficient to restrict the model completely. It is hard to figure out manually the under-constrained and over-constrained parts for such a large system. We use this case to show that the proposed method can assist the user to deal effectively with such a situation. Fig. 12 shows the two dependency vectors of the GCS for this case. From them, the computer can present to the user the two dependency groups: {C9,C16,C18,C20} and {C9,C25,C28,C29}. This isolation of the dependent constraints from the whole GCS can largely reduce the size of the constraints that the designer needs to consider in reasoning the constraint system. Consider, for example, the first dependency group. Cylinder F19 is parallel with plane F13 with constraint C16; plane F13 is also parallel to plane F14 with constraint C9; plane F14 is further parallel to plane F20 with constraint C20; therefore, cylinder F19 is parallel to plane F20. The remaining constraint C18 specifies the same parallelism between cylinder F19 and plane F20. Then, one can quickly understand the cyclical dependency among the constraints, and then make a decision of which constraint to remove. Therefore, such dependency groups are informative and useful to assist the designer to resolve redundancies in constraint systems.

Fig. 13 shows the result of applying Algorithm 1 to the third case. Eight maximal rigid subsystems were generated. With these subsystems in place, one only needs to focus on the constraints that bridge two subsystems for reasoning under-constraint. This isolation of the bridging constraints from the whole GCS also significantly reduces the constraints that the designer needs to consider in the under-constraint reasoning. For this case, there are only 14 bridging constraints (Fig. 14). What makes the reasoning much easier is that, for most subsystems pairs, there are only one or two bridging constraints, e.g., the subsystem pair P2,P8. Unfortunately, not all cases are of such a trivial saturation. The groups P1 and P2 have 5 bridging constraints. The reasoning of under-constraint for such a situation is more complicated but still manageable since the constraints that one needs to consider are small: without much reasoning, one can figure out that subsystem P2 has one rotational flexion motion and two translational flexion motions with respect to subsystem P1. Therefore, the information of the maximal rigid subsystems and bridging constraints is helpful for the designer to understand the under-constraint and specify additional constraints to resolve the under-constraint.

## 5. Conclusions

An approach to analyze variational B-rep models for direct modeling has been presented in this work. It features a generalization of the witness method so that it is able to directly handle variational B-rep models with the standard STEP representation scheme, and it is capable of correctly detecting under-, over-, well-constrained parts of a GCS and correctly decomposing the GCS. In addition, this approach can be readily extended to dealing with other schemes, as discussed in the last paragraph of Section 3.2. The developed methods have been evaluated through comparisons with the previous witness method as well as a series of case studies including a real-world mechanical design case. Some limitations of the proposed work are: as the focus of this work is geometric constraint analysis, the proposed method may be further from geometric constraint solving; it may complicate the equality test between geometric entities.

Some interesting future research directions for this work are summarized as follows. The present work focuses on the analysis of the geometric constraint system after direct modeling edits, giving information like over-constrained subsets. No changes are made to the geometric constraint system under analysis. The problems of adding new constraints and removing redundant constraints are to be studied in our next work. The proposed method, in the current form, can only deal with the constrained state with respect to translational/rotational transformations. Another useful transformation type – scaling – has not been included yet. It would be interesting to extend the results developed in this work for the scaling transformations, and the idea discussed in [23] may be used for this purpose. In addition, designing interactive interfaces for visualizing GCS analysis results is of great interest for future research studies.

### Acknowledgments

This research has been funded in part by the Natural Sciences and Engineering Research Council of Canada (NSERC) under the Discovery Grants program.

# Variational B-Rep Model Analysis

**Appendix: Proof of Proposition 1**

**Proof:** Rigid-body motions are evidently motions preserving model shape, and thus there is no need for a proof. The proof here is focused on geometry-invariant motions. Let $M$ be a model having $m$ geometric entities $\{g_i\}_{i=1}^m$. Without loss of generality, assume that there is a motion $T$ applied to the first $k$ geometric entities of the model, producing a perturbed model $M'$: $\{Tg_1, \ldots, Tg_k, g_{k+1}, \ldots, g_m\}$; and assume that $T$ preserves the model shape. As $M$ and $M'$ have the same shape, there exists a rigid-body transformation $T'$ that is able to transform $M$ to $M'$. This means that $T'g_i$ and $Tg_i$, $1 \leq i \leq k$, describe the same entity (they are geometrically same but the parameters may be different, e.g., two planes having collinear orientation vectors and coplanar positions are geometrically same). Therefore, the difference between the two transformations $T$ and $T'$ should be a geometry-invariant motion for the first $k$ geometry entities. This difference is to be denoted as $T_i''$, with $T' = T_i''T$. For the rigid-body transformation $T'$, its inverse $T'^{-1}$ is also a rigid-body transformation and can be used to transform $M'$ back to $M$, yielding $\{T'^{-1}Tg_1, \ldots, T'^{-1}Tg_k, T'^{-1}g_{k+1}, \ldots, T'^{-1}g_m\}$. Again, this means that $T'^{-1}g_i$ and $g_i$, $k+1 \leq i \leq m$, describe the same entity. Then, $T'^{-1}$ is not only a rigid-body motion but also a geometry-invariant motion for the geometric entities from $k+1$ to $m$. Note that if a transformation is geometry-invariant, its inverse is also a geometry-invariant motion. Thus, from the previous equation $T' = T_i''T$, we have that the applied motion $T = T_i''^{-1}T'$ is a composite of two geometry-invariant motions. If geometry-invariant motions are composited with rigid-body motions, the resulting motions obviously preserve the model shape, which concludes the proof. ∎